\documentclass[aps,prb,reprint,superscriptaddress,showpacs]{revtex4-1}

\usepackage{graphicx}
\usepackage[american]{babel}
\usepackage[utf8]{inputenc}
\usepackage[T1]{fontenc}
\usepackage{bbm,ae,aecompl}
\usepackage{amsmath}
\usepackage{amssymb}
\usepackage{amstext}
\usepackage{amsthm}
\usepackage{latexsym}
\usepackage{verbatim}
\usepackage{natbib}
\begin{document}

\title{Angle-resolved photoemission study of the role of nesting and orbital orderings in the antiferromagnetic phase of $\text{Ba}\text{Fe}_{2}\text{As}_{2}$}

\date{\today}

\author{M. Fuglsang Jensen}

\author{V. Brouet}

\author{E. Papalazarou}
\affiliation{Laboratoire de Physique des Solides, Universit\'{e} Paris-Sud, CNRS-UMR 8502, B\^at. 510, 91405 Orsay, France}

\author{A. Nicolaou}

\author{A. Taleb-Ibrahimi}

\author{P. Le F\'{e}vre}

\author{F. Bertran}
\affiliation{Synchrotron SOLEIL, L'Orme des Merisiers, Saint-Aubin-BP 48, 91192 Gif sur Yvette, France}

\author{A. Forget}

\author{D. Colson}
\affiliation{Service de Physique de l'Etat Condens\'{e}, Orme des Merisiers, CEA Saclay, CNRS-URA 2464, 91191 Gif sur Yvette Cedex, France}

\begin{abstract}
We present a detailed comparison of the electronic structure of BaFe$_2$As$_2$ in its paramagnetic and antiferromagnetic (AFM) phases, through angle-resolved photoemission studies. Using different experimental geometries, we resolve the full elliptic shape of the electron pockets, including parts of $d_{xy}$ symmetry along its major axis that are usually missing. This allows us to define precisely how the hole and electron pockets are nested and how the different orbitals evolve at the transition. We conclude that the imperfect nesting between hole and electron pockets explains rather well the formation of gaps and residual metallic droplets in the AFM phase, provided the relative parity of the different bands is taken into account. Beyond this nesting picture, we observe shifts and splittings of numerous bands at the transition. We show that the splittings are surface sensitive and probably not a reliable signature of the magnetic order. On the other hand, the shifts indicate a significant  redistribution of the orbital occupations at the transition, especially within the $d_{xz}$/$d_{yz}$ system, which we discuss. \end{abstract}

\maketitle

\section{Introduction}
Since the recent discovery of iron-based superconductors, the relationship between magnetism and superconductivity has been a central question \cite{Paglione:2010p271}. A better understanding of the nature of the magnetic state is needed to go further. In BaFe$_2$As$_2$, the high temperature phase is tetragonal and paramagnetic (PM) and transforms into an orthorhombic and antiferromagnetic (AFM) phase below 137 K \cite{Rotter:2008p253}. Neutron scattering has shown that the magnetic moments on Fe align co-linearly; antiferromagnetically along $x$ and ferromagnetically along $y$ (we call $x$ and $y$ the Fe first nearest neighbor directions) \cite{Huang:2008p252}. As the AFM phase is still metallic, it seems natural to describe it within an itinerant spin density wave (SDW) picture. The fact that the AFM wave vector corresponds to a good nesting vector between the hole and electron pockets indeed first suggested that a classical SDW picture could apply \cite{Mazin:2008p254}. However, the subsequent discovery of a distinct AFM ordering in FeTe, despite an essentially similar Fermi surface (FS), complicated the picture and cast doubts that nesting could be the only driving force for the transition \cite{Johannes:2009p255}. Very early, a local moment picture was also advocated \cite{Yildirim:2008p257} and the essential role of Hund's rule coupling in the formation of the local moments receives more and more attention \cite{Yin:2010p228}. This coupling is a consequence of the multiband nature of these compounds and the role of orbital orderings at, or before, the magnetic transition, is indeed actively discussed \cite{Lv:2009p260,Lee:2009}.

Angle-resolved photoelectron spectroscopy (ARPES) is a very well suited probe of a SDW transition, as it can be used to image the FS, identify the best nested parts and relate this to the size of gaps at different FS locations \cite{Brouet:2008}. Furthermore, in a complex system like iron pnictides, it can resolve the properties of the different hole and electron bands and give valuable clues on their orbital origin \cite{Zhang:2009,Mansart:2010}. In BaFe$_2$As$_2$, the changes at the AFM transition first appeared subtle \cite{Fink:2009p30,ZabolotnyyNature09}, but a number of significant features have now been identified by ARPES. The hole pockets at $\Gamma$ lose their circular shape and become \textit{flower-like}, where each petal is in fact a tiny electron pocket \cite{Hsieh:2008p64}. The electron pockets, located at the corners of the PM Brillouin zone (BZ), split into 4 \textit{droplets} \cite{Jong:2010p95,LiuZhouPRB09,LiuKaminskiNaturePhys10}, later described as Dirac cones \cite{Richard:2010p62}. At the same time, new bands, apparently folded with the magnetic periodicity, clearly appear \cite{Hsieh:2008p64,Jong:2010p95,Kondo:2010p261}. Although such residual metallic pockets could be expected in a SDW with imperfect nesting \cite{Brouet:2008}, their relationship with the PM electronic structure has not been clarified \cite{LiuZhouPRB09}. In addition, a curious observation is a splitting of some bands at the transition that was attributed either to an exotic form of exchange splitting \cite{Yang:2009p262} or to an anisotropy appearing between the $x$ and $y$ direction observed simultaneously in twinned samples \cite{Hsieh:2008p64,Yi:2009p234}. More recently, experiments in de-twinned samples have appeared \cite{KimPRB11,YiPNAS11,DessauUntwinned}, but they did not solve this issue. Finally, orbital splitting between $d_{xz}$ and $d_{yz}$ was claimed from a 2-fold symmetry of the hole FS \cite{Shimojima:2010p37}, which is a somewhat puzzling result, as most reported hole FS in fact appear 4-fold \cite{Hsieh:2008p64,Yi:2009p234,Kondo:2010p261}. Obviously, there is also an important contribution of other orbitals than $d_{xz}$ and $d_{yz}$ to the electronic structure \cite{Zhang:2009,Mansart:2010} and their respective role in the transition is still quite unclear. 

Recently, we have obtained a better understanding of the structure of the electron pockets, by finding an experimental condition where the electron bands along the major and minor axis of the elliptical electron pocket can be clearly isolated \cite{BrouetJensen2011}. This is useful for the study of the magnetic transition in at least two aspects. First, this allows to define precisely the shape of the electron pocket and therefore the real degree of nesting between the hole and electron pockets. We discuss the relevance of these nesting ideas to describe the reconstruction of the electronic structure in the SDW state. Second, these bands have very anisotropic dispersions depending on their orbital character ($d_{xz}$/$d_{yz}$ or $d_{xy}$), which allows to follow clearly what happens to each band at the transition. This is impossible for the hole bands, which all have similar dispersions and quite strongly overlap. In particular, we stress the role of the $d_{xy}$ band in the formation of the \lq\lq{}Dirac cone\rq\rq{} and we reinterpret one \lq\lq{}folded hole band\rq\rq{} as a $d_{yz}$ band shifted up, in good agreement with a recent study in de-twinned sample \cite{YiPNAS11}. In addition, we observe that the splitting of the bands at the transition depends on the surface properties and may not be relevant for the understanding of magnetism. We then estimate the degree of polarization in the $d_{xz}$/$d_{yz}$ orbitals. 

 BaFe$_2$As$_2$ single crystals were grown using a FeAs self-flux method.  ARPES experiments were carried out at the CASSIOPEE beamline at the SOLEIL synchrotron, with a Scienta R4000 analyser, an angular resolution of $0.2^{\circ}$ and an energy resolution better than $15$ meV. 

\section{Paramagnetic phase}
\begin{figure}[tbp]
\centering
\includegraphics[width=0.5\textwidth]{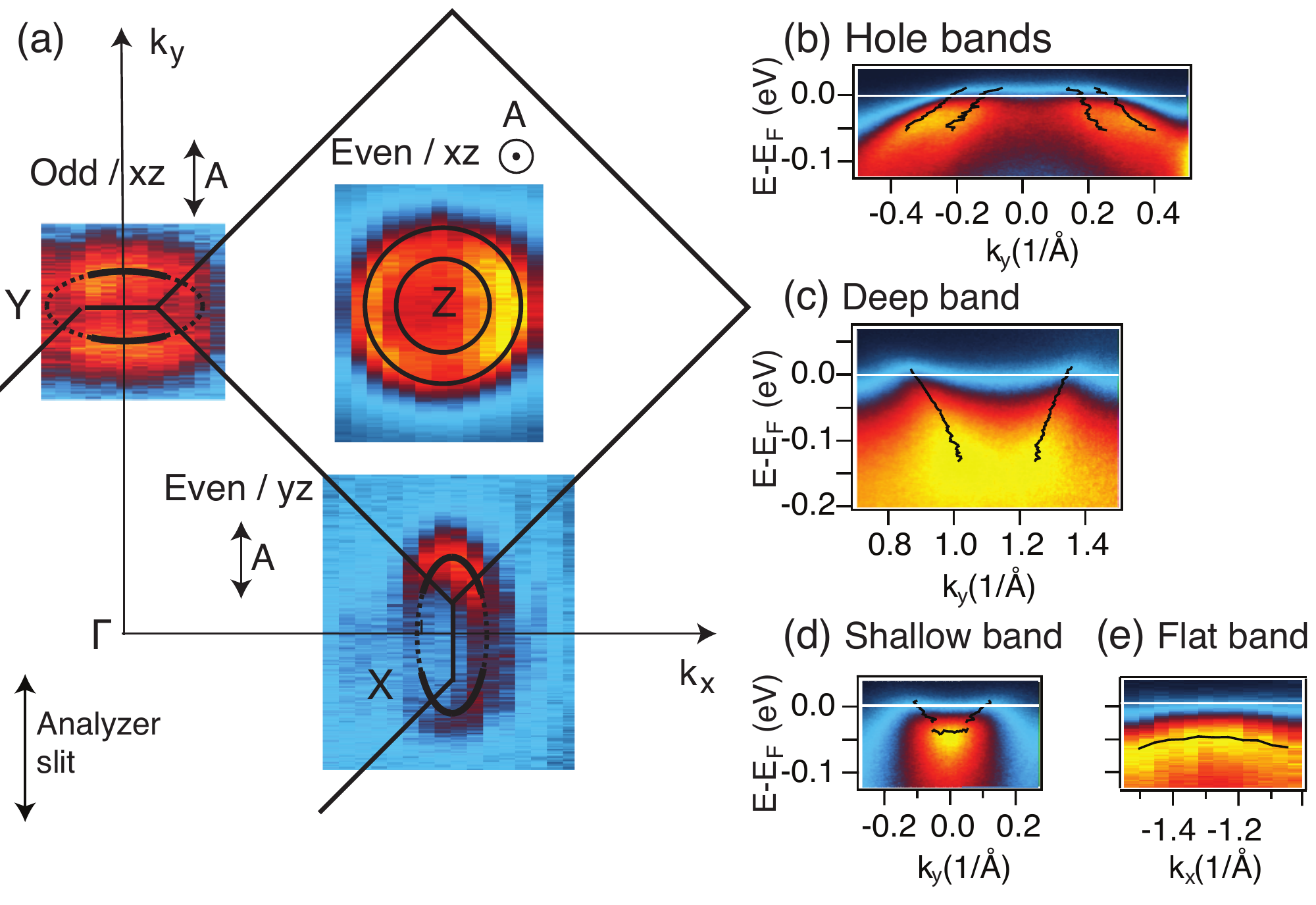}
\caption{(a) FS measured in the PM state at T=150 K with $\hbar \omega=34$ eV photons, obtained by integration of the spectral intensity in a 10meV window around $E_F$. The direction of the light polarization, $\mathbf{A}$, is indicated on the map for each measurement. The plane of light incidence is $xz$ and the analyser slits are fixed along $k_y$. (b-e) Energy-momentum intensity plots along (b) $k_{x}$ through Z, (c) $k_{y}$ through X, (d) $k_{y}$ through Y, (e) $k_{x}$ through Y. Black lines indicate dispersions determined experimentally. X and Y are equivalent points in the BZ, but are given different names here for clarity.}
\label{PM-figure}
\end{figure}
\subsection{Fermi Surface}
In Fig.\ \ref{PM-figure}(a), the FS is presented at $T=150$~K in the PM state. The photon energy is chosen such that the normal emission corresponds to the Z point of the BZ  \cite{Brouet:2009p106}. Two concentric and circular hole pockets are observed around the Z point with diameters $2k_{F}=0.2(2)$ and $0.47(2)$~\AA$^{-1}$ (Fig.\ \ref{PM-figure}(b)). The inner one is believed to be doubly degenerate and formed by the $d_{xz}$ and $d_{yz}$ orbitals, while the outer one is not degenerate and of even symmetry \cite{Zhang:2009,Mansart:2010}. The electron pockets are measured in two different experimental geometries, which select orbitals either \textit{even} with respect to the $yz$ plane or \textit{odd} with respect to the $xz$ plane. Note that the even parts are missing in most ARPES measurements, leaving it unclear how the electron pockets close towards Z. More details about how this band is detected can be found in ref.\cite{BrouetJensen2011}. At $k_z=1$, the electron pocket is elliptic and oriented towards Z, odd along its minor axis ($2k_{F}=0.18(4)$ \AA$^{-1}$) and even along its major axis ($2k_{F}=0.54(2)$ \AA$^{-1}$). The dispersions of the bands are very different; the odd parts form a very shallow band of about 40 meV (Fig.\ \ref{PM-figure}(d)), while the even band is deeper (100 meV, see Fig.\ \ref{PM-figure}(c)). The shallow band corresponds very well to the odd $d_{xz}$/$d_{yz}$ in theoretical calculations and the deep band to $d_{xy}$, although its parity is not expected as predominantly even (see Fig.\ 4 or ref.\cite{BrouetJensen2011}). The $d_{xz}$/$d_{yz}$ band also forms a flat band below $E_F$ along the ellipse major axis, shown in Fig.\ \ref{PM-figure}(e). A second electron pocket is expected at each BZ corner, arising from folding in the reduced BZ \cite{Graser:2010p67}. We have shown \cite{BrouetJensen2011} that it is suppressed in these experimental conditions, which simplifies greatly the structure and will allow to better understand the modifications in the AFM state. 

\subsection{Quality of Nesting}
\begin{figure}[tbp]
\centering
\includegraphics[width=0.5\textwidth]{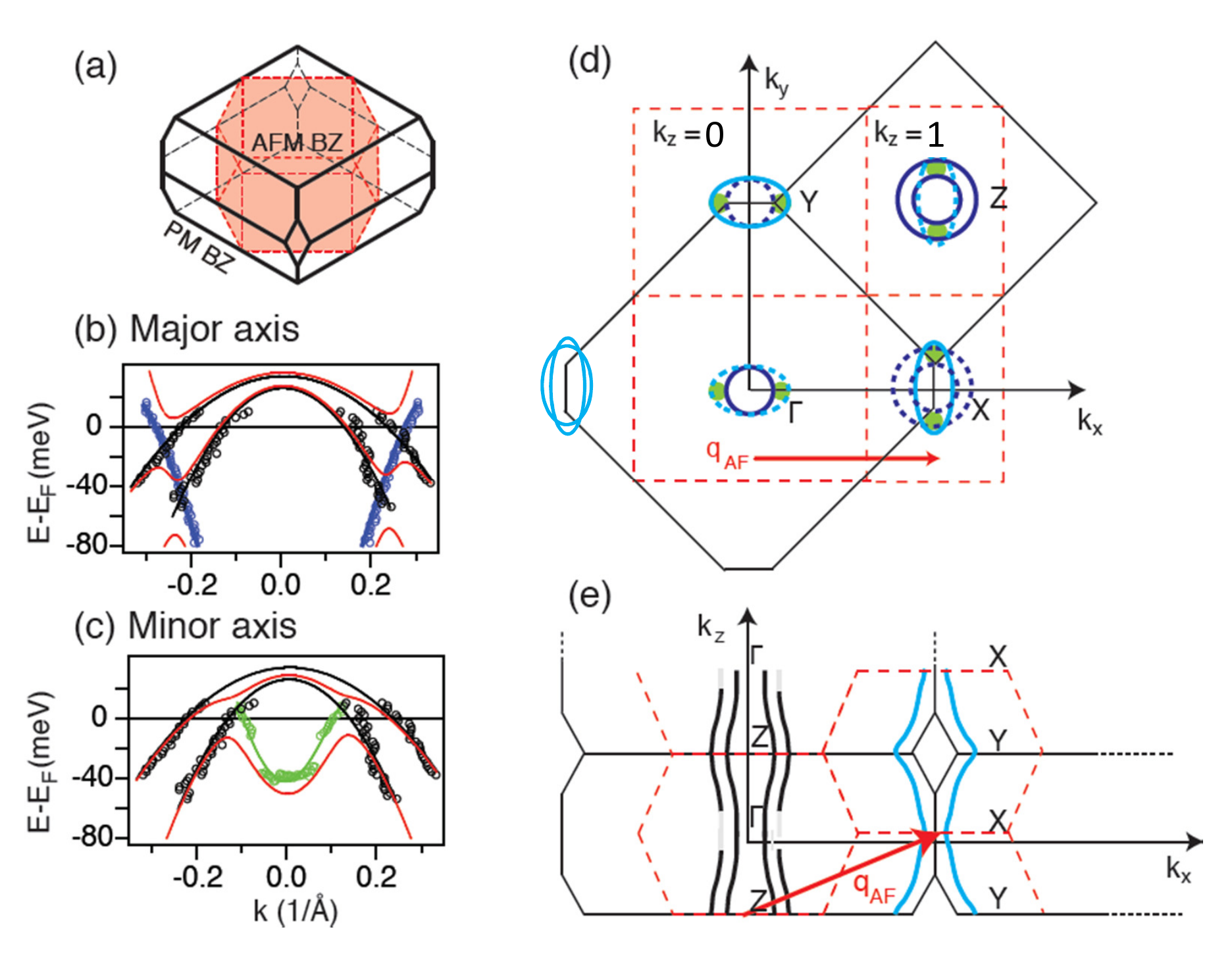}
\caption{(a) Sketch of the 3D AFM and PM BZ. (b-c) Dispersions measured at $k_{z}=1$ in the PM phase for hole and electron bands, along the major (b) and minor axis (c). Red lines sketch the dispersion expected after interaction between the electron and hole bands. (d) Sketch of the PM FS (thick solid lines) with parts translated with $\mathbf{q}_{AF}$ as dashed lines. Thin red dashed lines indicate the AFM BZ boundaries. The structurally folded electron pockets are omitted except on the left corner. (e) Sketch of the BZ in $xz$ plane. Thick lines indicate hole and electron pockets.}
\label{nestingfig}
\end{figure}

We now consider which electronic structure would be expected in the magnetic state within a simple SDW picture. Fig.\ \ref{nestingfig}(a) shows how the 3D magnetic BZ compares with the PM BZ. The stacking of the magnetically ordered FeAs planes is such \cite{Huang:2008p252} that neighboring magnetic BZ along the AFM direction $k_x$ are shifted along $k_z$ (Fig.\ \ref{nestingfig}(e)). In Fig.\ \ref{nestingfig}(d), we sketch the FS observed in the PM state by solid lines and we show by dashed lines the pockets translated by the AFM wave vector, $\mathbf{q}_{AF}$ [note that its has a $k_z$ component, see Fig.\ \ref{nestingfig}(e). At $k_z=1$, the sizes of the circles and ellipses were determined experimentally. At $k_z=0$, the hole pockets are smaller and the outer pocket is not clearly detected, as sketched in Fig.\ \ref{nestingfig}(e), so that we only represent the smaller circle. The electron pocket essentially rotates with $k_z$, but also becomes more rounded as sketched for $k_z=0$ \cite{Graser:2010p67,BrouetJensen2011}. We omit for clarity, and because it is not detected experimentally, the structurally folded electron pocket, except on the left corner (this folding ensures the equivalency between X and Y). 

Due to the elliptic shape of the electron pocket, the nesting is very different along the minor and major ellipse axis. Along the major axis (Fig.\ \ref{nestingfig}(b)), the deep electron band crosses the inner hole band and is relatively well nested with the outer hole band, while, along the minor axis (Fig.\ \ref{nestingfig}(c)), it is the shallow electron band that is now nested with the inner hole band. Assuming the interaction between the original and folded bands will open a gap of 40meV at their crossings, we obtain the dispersions sketched as red lines \cite{Brouet:2008}. An electron-like droplet should appear between the inner and outer hole bands, with a gap on the outer side. These droplets are shown as green filled circles on the FS map. On the other hand, the shallow electron band should be completely gapped. The details of the droplets and gaps should change along $k_z$, because the sizes of the pockets slightly change. This can actually be observed \cite{ZhouFengPRB10}, but we neglect such effects here, as we have found that they do not modify the general picture. 

\section{Magnetic phase}
\begin{figure}[tbp]
\centering
\includegraphics[width=0.48\textwidth]{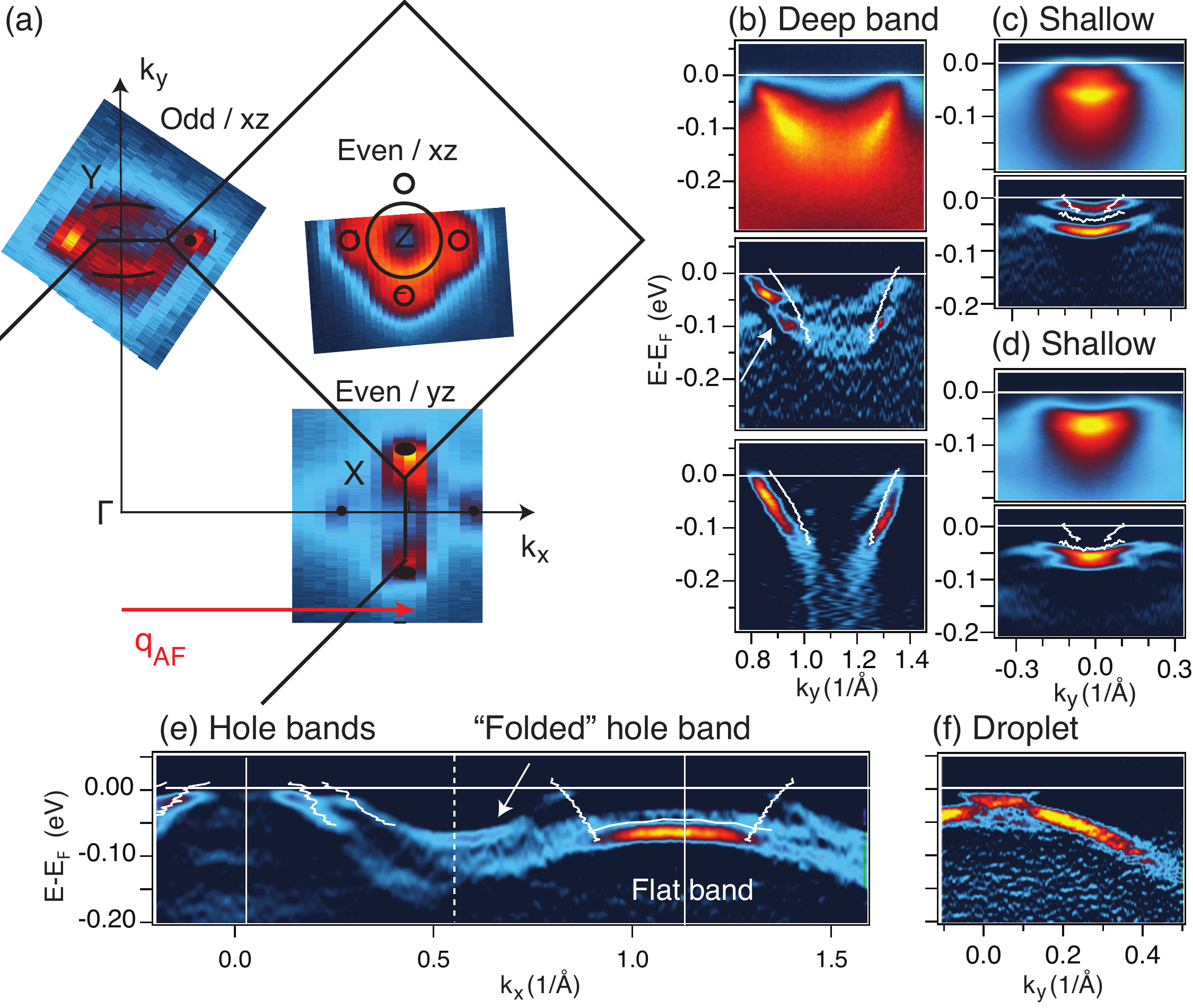}
\caption{(a) FS in the AFM state at T=20 K and with the same conditions as in Fig.\ \ref{PM-figure}. (b) Energy-momentum intensity plot of the even electron band along $k_{y}$ through X, as well as second derivative (SD) image with respect to energy (center) and k (lowest). (c-d) Energy-momentum intensity plots and SD images with respect to k of the shallow electron band along $k_{y}$ through Y in two different surface conditions. (e) SD along $k_{x}$ through Z in the odd geometry. (f) SD through a droplet near the Z-point in the $k_{x}$ direction.}
\label{AFM-figure}
\end{figure}

In Fig.\ \ref{AFM-figure}, we present the FS in the AFM phase at $T=20$~K, together with dispersions of the main features. The PM dispersions are added for reference to the second derivative images as white lines. From the nesting picture described just before, we would expect that \textit{gaps} and \textit{droplets} appear near the Fermi level, but that the electronic structure remains essentially the same above the energy scale of these gaps. Indeed, there is at first sight not much change on the position of the hole bands (Fig.\ \ref{AFM-figure}(e)). On the other hand, all electron bands (deep, shallow and flat) seem to \textit{shift} downwards by about 20meV and sometimes \textit{split} (especially the shallow one in Fig.\ \ref{AFM-figure}(c)) This indicates a larger reorganization of the electronic structure. Except for the behavior of the deep electron band, similar features were reported before, as recalled in the introduction. However, they were rarely discussed altogether, so that a comprehensive picture of the complex modifications of the electronic structure at the magnetic transition is still missing. For example, the center and corners of the PM BZ become equivalent in the AFM BZ (except for the different $k_z$), so why do they look so different? Is it just due to differences in relative intensities of hole and electron bands at the two points or are there different structures (e.g. droplets and Dirac cones)? Why would the electron bands shift and not the hole bands? How is the number of carriers conserved if electron bands just shift down? Do the splitting and shifting originate from the same interaction ? In the following, we address all these questions by reviewing successively the different types of modifications appearing at the transition. 

\subsection{Residual metallic droplets}

A clear modification is the appearance of 4 electron droplets on the outskirt of the hole pockets, marked by small circles on the FS \cite{Hsieh:2008p64,Yi:2009p234}. They are located between the inner and outer hole pockets (Fig.\ \ref{AFM-figure}(e)) and are more clearly seen in Fig.\ \ref{AFM-figure}(f) that presents a cut through such a droplet. They correspond quite well to the expected green droplets in Fig.\ \ref{nestingfig}(d), because twinning creates a 4-fold symmetry. They can therefore be assigned to the interaction between the inner hole band and the deep electron band. It has rarely been noticed, but it is clear from Fig.\ \ref{nestingfig}, that twinning should \textit{not} produce a 4-fold symmetry at the electron pocket (this point is often missed, because the second electron pocket, arising from structural folding \cite{Graser:2010p67,BrouetJensen2011} is often incorrectly represented rotated by 90$^\circ$). The 4 bright spots appearing at the extremities of the electron pockets and marked by black points on the FS must then have a different origin. In fact, the interaction forming the green droplet is also visible on the deep electron band. The white arrow in Fig.\ \ref{AFM-figure}(b) indicates the crossing with the folded inner hole band, but it is less clear, as the folded inner hole band is very weak. Most of the intensity of the bright spots is located where the deep electron band crosses $E_F$. 

Fig.\ \ref{AFM-figure} allows us to understand that the bright spots are in fact created by different bands in the different cases (some residual intensity due to the shallow electron band, which sometimes look point-like, is also observed in the odd measurement). The spots along $k_{y}$ in the even measurement are due to the deep electron band, while those along $k_x$, both in odd and even measurements, are due to a \lq\lq{}folded\rq\rq{} hole band, as the one shown in Fig.\ \ref{AFM-figure}(e). Indeed, this band looks like the outer hole band folded with respect to the AFM BZ (dashed line), although it has a surprisingly strong intensity for a folded band. We will return to its exact nature later. As the deep electron band is usually not observed, most of the spots reported so far are in fact due to the folded hole band and seem to appear outside the contour of the PM FS \cite{Jong:2010p95,Richard:2010p62,LiuZhouPRB09}, which seems rather strange. Fig.\ \ref{AFM-figure}(e) makes it clear that they are in fact located where the folded band meets the deep electron band, forming a sort of \lq\lq{}Dirac cone\rq\rq{} in this direction. However, the two bands are never observed together in our measurements, evidencing they have opposite parity. 

\subsection{SDW Gaps}

\begin{figure}[tbp]
\centering
\includegraphics[width=0.48\textwidth]{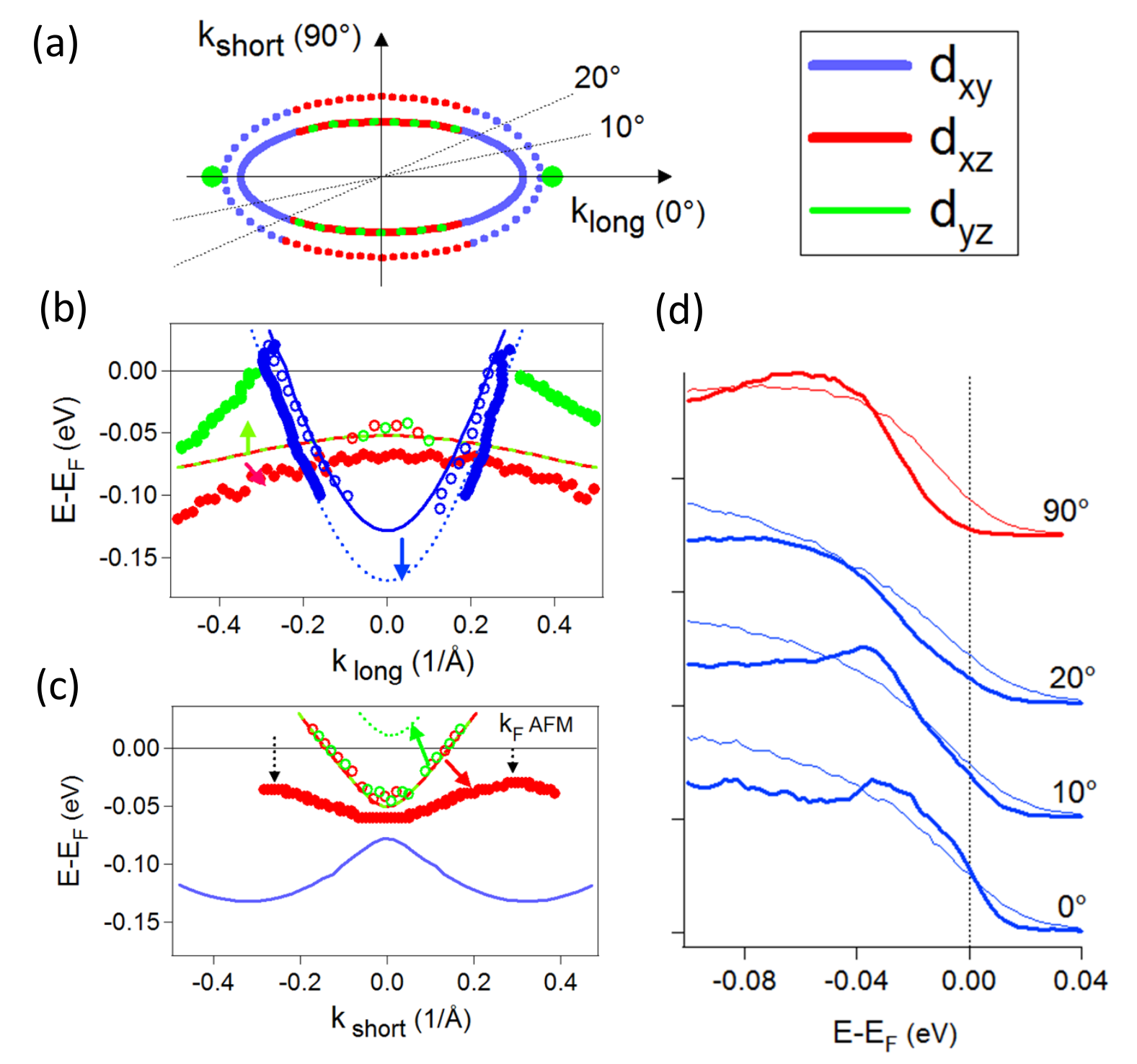}
\caption{(a) Sketch of the electron ellipse in the PM phase (thick line) and in the AFM phase (dotted line). Colors indicate the orbital character. (b-c) Dispersion in the PM phase (open symbols) and the AFM phase (closed symbols) along the ellipse major axis (b) and minor axis (c). Thick lines indicate the dispersions calculated in the PM phase and renormalized by a factor 3 \cite{BrouetJensen2011}.  (d) Spectra at $k_F$ at 150K (thin lines) and 20K (thick lines) for different cuts of the ellipse identified by their with respect to the ellipse major axis. }
\label{AFM-shift}
\end{figure}

Fig.\ \ref{AFM-shift}(d) presents the gaps opening in the AFM phase, by comparing spectra measured at $k_F$ along different cuts of the electron ellipse, at 20 and 150 K. There is clearly no gap on the deep electron band along the major axis (bottom spectra), despite its good nesting conditions with the outer hole band. This is likely due to the opposite parity of the deep and folded hole bands, which we have just evidenced experimentally and which should forbid their interaction. Away from $k_{y}$, the deep electron band acquires some odd character and becomes quickly gapped, as shown by the much smaller extension of the FS in the AFM state compared to PM one and directly by the spectra of Fig.\ \ref{AFM-shift}(d). Such a symmetry argument was used by Ran et al.\cite{Ran:2009p157} to predict the formation of Dirac cones at these particular points. Parity rather than nesting is essential to describe these features. As for the shallow electron band, we do observe that it is almost entirely gapped (top spectra in Fig.\ \ref{AFM-shift}(d)), in agreement with Fig.\ \ref{nestingfig}(c), although the situation is complicated there by the splitting, evidently changing the nesting conditions. 

A revised nesting picture, taking into account symmetry dependent interactions, then describes well the formation of gaps and droplets in this phase. The order of magnitude of these interactions is 20-40 meV, which agrees well with the smallest gap detected in optical measurements \cite{moon:2010}. 

\subsection{Splittings and surface effects}

The splitting of the shallow band, as well as that of some other bands, was reported before as an unusual feature of this state \cite{Yang:2009p262,Yi:2009p234}. We reveal here that it depends on the surface structure. Among 10 samples cleaved at low temperatures, we observed in half the cases a clear splitting of 40~ meV as in Fig.\ \ref{AFM-figure}(c) and in the other half, a much smaller splitting of 15~meV as in Fig.\ \ref{AFM-figure}(d). Moreover, we observed that annealing a sample with a (d)-like splitting up to 300~K and recooling it below the transition, results in a (c)-like splitting. STM and LEED studies have shown that, when a sample is cleaved cold, the Ba remaining at the surface may adopt particular orders, which are irreversibly lost after annealing at 300 K \cite{Massee:2009p247}. It is therefore very likely that the splitting depends on the Ba order and is not intrinsic to the magnetic state. On the other hand, the downward shift of the lower band by about 20~meV in both cases seems to be a robust feature and was for example also observed in NaFeAs, which has quite a different surface \cite{He:2010p245}. 

It would be interesting to clarify the origin of this splitting to understand how the structure may influence the magnetism and also to what extent these ARPES data are representative of the bulk magnetic structure. The fact that similar gaps are measured by ARPES and other techniques supports the idea that ARPES still probes essentially the bulk magnetic structure. At present, we could either attribute the lower band to a bulk band and the upper band to a surface band or suppose that there is some small ferromagnetic component at the surface, which splits the bands and whose magnitude depends on the surface structure. 
 
\subsection{Shifts and orbital ordering}
The most original feature of the magnetic state may be the shifts observed at the transition. As shifting a band relatively to $E_{F}$ changes the number of carriers it contains, this implies a redistribution of the occupation of the different orbitals at the transition and is a signature of the multi-orbital nature of the transition. Qualitatively, it seems natural that the AFM or ferromagnetic alignment of the spins along $x$ or $y$ affects the orbital occupations in the two directions. Our ARPES data allow some quantitative estimates. Fig.\ \ref{AFM-shift}(a) sketches the increase of the size of the pocket in the AFM phase, corresponding to the down shifts of the electron bands we observe. As recalled in Fig.\ \ref{AFM-shift}(b), the deep electron band shifts down by 30 meV, increasing its diameter to $2k_{F}=0.60(2)$~\AA$^{-1}$ and the lower shallow band (Fig.\ \ref{AFM-shift}(c)) by 20~meV, increasing its diameter to $2k_{F}=0.26(4)$ \AA$^{-1}$ ($k_{F}$ is defined as the point closest to $E_{F}$, even if it is gapped, see Fig.\ \ref{AFM-shift}(c)). The number $n$ of carriers contained in one pocket is simply proportional to the FS area A$_{FS}$ ($n$=2A$_{FS}$/A$_{BZ}$ \cite{Brouet:2009p106}). This increases from 0.035 electrons/Fe in the PM state for each electron pocket to 0.05 electrons/Fe in the AFM state. This rough estimate suggests that one pocket must be nearly empty in the AFM state, indicating a significant orbital polarization. As the $d_{xz}$/$d_{yz}$ band is quite shallow, one would need to shift it up only by 50~meV to empty it completely. If this happens, this band will not be observed anymore with ARPES, but a similar shift should occur on the flat electron band, whose top coincides with the bottom of the shallow band. Fig.\ \ref{AFM-shift}(b) shows that it is indeed possible to interpret the folded hole band as the flat band shifted up. Its curvature seems larger than that of the bottom band, but this is a very simple model and this bands probably also interacts with the folded hole bands expected in this region. This would explain the strong intensity of this band and also its polarization behavior, which is not that of the outer hole band. In Fig.\ \ref{AFM-figure}, it appears both even with respect to $yz$ and odd with respect to $xz$. Only $d_{xz}$ or $d_{yz}$ (but not both) can produce this behavior, even in the presence of twinning (the same argument was used in \cite{Shimojima:2010p37} for a different situation). In fact, this is in nice agreement with a recent experiment in a de-twinned sample \cite{YiPNAS11}, where it is observed that the flat electron band shifts down along $x$ and up along $y$, creating the \lq\lq{}folded hole band\rq\rq{}. We also note that in these detwinned experiments \cite{YiPNAS11,KimPRB11}, both bands were split, indicating that the splitting is something different, not associated with twinning. This is consistent with our observation that it is surface dependent. This also means that what looks like the outer hole band in Fig.\ \ref{AFM-figure}(e) is in fact more likely the inner $d_{yz}$ band shifted up. Therefore there are probably similar shifts in the hole bands, but they are more difficult to decipher as the bands strongly overlap and all have similar dispersions. Finally, these estimations validate the sketch of Fig.\ \ref{AFM-shift}(a) with a large electron pocket along $y$ and only small residual pockets along $x$, probably near the $d_{xy}$ parts. 

\section{Conclusion} 

To conclude, we have shown that the FS nesting explains well the formation of gaps and residual metallic pockets, provided that the parity of the different bands is taken into account. However, the gaps detected here are rather small (at most 40~meV) and are of the same order of magnitude as energy shifts detected for many bands. This means that the gain of energy in the AFM state cannot simply be assigned to gap openings at $E_{F}$, as in a simple SDW case, but that reordering between different orbitals is as important. We stress that we here focus on the near $E_{F}$ region, whereas the higher energy scale may be quite different \cite{moon:2010} and also hold important keys to stabilize the magnetic state \cite{Yin:2010p228}. We show that the splitting of the $d_{xz}$/$d_{yz}$ band appearing at the magnetic transition \cite{Yang:2009p262} depends on the surface structure. However, we clearly identify a robust down shift of a band of $d_{xz}$/$d_{yz}$ symmetry (the shallow electron band) and an up shift of a band of opposite $d_{xz}$/$d_{yz}$ symmetry (the "folded" hole band). This establishes a significant degree of orbital polarization in the $d_{xz}$/$d_{yz}$ system that could be compared with different models of orbital orderings at the transition \cite{Lv:2009p260,Lee:2009}.  However, we also observe that other orbitals play an important role and move at the transition, most notably the deep electron band of $d_{xy}$ symmetry forming the extremities of the electron ellipse. These parts remain largely ungapped in the AFM state and may be the first parts where superconductivity will develop.

\begin{acknowledgments} 
We thank M. Aichhorn and S. Biermann for useful discussions and providing the LDA calculations. Financial support from the French RTRA Triangle de la physique and the ANR \lq\lq{}Pnictides\rq\rq{} is acknowledged. 
\end{acknowledgments}

\bibliography{finalbib2}

\begin{thebibliography}{10}%
\makeatletter
\providecommand \@ifxundefined [1]{%
 \ifx #1\undefined \expandafter \@firstoftwo
 \else \expandafter \@secondoftwo
\fi
}%
\providecommand \@ifnum [1]{%
 \ifnum #1\expandafter \@firstoftwo
 \else \expandafter \@secondoftwo
\fi
}%
\providecommand \enquote [1]{``#1''}%
\providecommand \bibnamefont  [1]{#1}%
\providecommand \bibfnamefont [1]{#1}%
\providecommand \citenamefont [1]{#1}%
\providecommand\href[0]{\@sanitize\@href}%
\providecommand\@href[1]{\endgroup\@@startlink{#1}\endgroup\@@href}%
\providecommand\@@href[1]{#1\@@endlink}%
\providecommand \@sanitize [0]{\begingroup\catcode`\&12\catcode`\#12\relax}%
\@ifxundefined \pdfoutput {\@firstoftwo}{%
 \@ifnum{\z@=\pdfoutput}{\@firstoftwo}{\@secondoftwo}%
}{%
 \providecommand\@@startlink[1]{\leavevmode\special{html:<a href="#1">}}%
 \providecommand\@@endlink[0]{\special{html:</a>}}%
}{%
 \providecommand\@@startlink[1]{%
  \leavevmode
  \pdfstartlink
   attr{/Border[0 0 1 ]/H/I/C[0 1 1]}%
   user{/Subtype/Link/A<</Type/Action/S/URI/URI(#1)>>}%
  \relax
 }%
 \providecommand\@@endlink[0]{\pdfendlink}%
}%
\providecommand \url  [0]{\begingroup\@sanitize \@url }%
\providecommand \@url [1]{\endgroup\@href {#1}{\urlprefix}}%
\providecommand \urlprefix [0]{URL }%
\providecommand \Eprint[0]{\href }%
\@ifxundefined \urlstyle {%
  \providecommand \doi [1]{doi:\discretionary{}{}{}#1}%
}{%
  \providecommand \doi [0]{doi:\discretionary{}{}{}\begingroup
  \urlstyle{rm}\Url }%
}%
\providecommand \doibase [0]{http://dx.doi.org/}%
\providecommand \Doi[1]{\href{\doibase#1}}%
\providecommand \bibAnnote [3]{%
  \BibitemShut{#1}%
  \begin{quotation}\noindent
    \textsc{Key:}\ #2\\\textsc{Annotation:}\ #3%
  \end{quotation}%
}%
\providecommand \bibAnnoteFile [2]{%
  \IfFileExists{#2}{\bibAnnote {#1} {#2} {\input{#2}}}{}%
}%
\providecommand \typeout [0]{\immediate \write \m@ne }%
\providecommand \selectlanguage [0]{\@gobble}%
\providecommand \bibinfo [0]{\@secondoftwo}%
\providecommand \bibfield [0]{\@secondoftwo}%
\providecommand \translation [1]{[#1]}%
\providecommand \BibitemOpen[0]{}%
\providecommand \bibitemStop [0]{}%
\providecommand \bibitemNoStop [0]{.\EOS\space}%
\providecommand \EOS [0]{\spacefactor3000\relax}%
\providecommand \BibitemShut [1]{\csname bibitem#1\endcsname}%
\bibitem{Paglione:2010p271}%
  \BibitemOpen
  \bibfield{author}{%
  \bibinfo {author} {\bibfnamefont{J.}~\bibnamefont{Paglione}}\ and\ \bibinfo
  {author} {\bibfnamefont{R.~L.}\ \bibnamefont{Greene}},\ }%
  \bibfield{journal}{%
  \bibinfo {journal} {Nature Phys.}\ }%
  \textbf{\bibinfo {volume} {6}},\ \bibinfo {pages} {645} (\bibinfo {year}
  {2010})%
  \bibAnnoteFile{NoStop}{Paglione:2010p271}%
\bibitem{Rotter:2008p253}%
  \BibitemOpen
  \bibfield{author}{%
  \bibinfo {author} {\bibfnamefont{M.}~\bibnamefont{Rotter}} \emph{et~al.},\ }%
  \bibfield{journal}{%
  \bibinfo {journal} {Phys. Rev. B}\ }%
  \textbf{\bibinfo {volume} {78}},\ \bibinfo {pages} {020503} (\bibinfo {year}
  {2008})%
  \bibAnnoteFile{NoStop}{Rotter:2008p253}%
\bibitem{Huang:2008p252}%
  \BibitemOpen
  \bibfield{author}{%
  \bibinfo {author} {\bibfnamefont{Q.}~\bibnamefont{Huang}} \emph{et~al.},\ }%
  \bibfield{journal}{%
  \bibinfo {journal} {Phys. Rev. Lett.}\ }%
  \textbf{\bibinfo {volume} {101}},\ \bibinfo {pages} {257003} (\bibinfo {year}
  {2008})%
  \bibAnnoteFile{NoStop}{Huang:2008p252}%
\bibitem{Mazin:2008p254}%
  \BibitemOpen
  \bibfield{author}{%
  \bibinfo {author} {\bibfnamefont{I.~I.}\ \bibnamefont{Mazin}}, \bibinfo
  {author} {\bibfnamefont{D.~J.}\ \bibnamefont{Singh}}, \bibinfo {author}
  {\bibfnamefont{M.~D.}\ \bibnamefont{Johannes}},\ and\ \bibinfo {author}
  {\bibfnamefont{M.~H.}\ \bibnamefont{Du}},\ }%
  \bibfield{journal}{%
  \bibinfo {journal} {Phys. Rev. Lett.}\ }%
  \textbf{\bibinfo {volume} {101}},\ \bibinfo {pages} {057003} (\bibinfo {year}
  {2008})%
  \bibAnnoteFile{NoStop}{Mazin:2008p254}%
\bibitem{Johannes:2009p255}%
  \BibitemOpen
  \bibfield{author}{%
  \bibinfo {author} {\bibfnamefont{M.~D.}\ \bibnamefont{Johannes}}\ and\
  \bibinfo {author} {\bibfnamefont{I.~I.}\ \bibnamefont{Mazin}},\ }%
  \bibfield{journal}{%
  \bibinfo {journal} {Phys. Rev. B}\ }%
  \textbf{\bibinfo {volume} {79}},\ \bibinfo {pages} {220510} (\bibinfo {year}
  {2009})%
  \bibAnnoteFile{NoStop}{Johannes:2009p255}%
\bibitem{Yildirim:2008p257}%
  \BibitemOpen
  \bibfield{author}{%
  \bibinfo {author} {\bibfnamefont{T.}~\bibnamefont{Yildirim}},\ }%
  \bibfield{journal}{%
  \bibinfo {journal} {Phys. Rev. Lett.}\ }%
  \textbf{\bibinfo {volume} {101}},\ \bibinfo {pages} {057010} (\bibinfo {year}
  {2008})%
  \bibAnnoteFile{NoStop}{Yildirim:2008p257}%
\bibitem{Yin:2010p228}%
  \BibitemOpen
  \bibfield{author}{%
  \bibinfo {author} {\bibfnamefont{Z.~P.}\ \bibnamefont{Yin}}, \bibinfo
  {author} {\bibfnamefont{K.}~\bibnamefont{Haule}},\ and\ \bibinfo {author}
  {\bibfnamefont{G.}~\bibnamefont{Kotliar}},\ }%
  \bibfield{journal}{%
  \bibinfo {journal} {Nature Physics}\ }%
  \textbf{\bibinfo {volume} {7}},\ \bibinfo {pages} {294} (\bibinfo {year}
  {2011})%
  \bibAnnoteFile{NoStop}{Yin:2010p228}%
\bibitem{Lv:2009p260}%
  \BibitemOpen
  \bibfield{author}{%
  \bibinfo {author} {\bibfnamefont{W.}~\bibnamefont{Lv}}, \bibinfo {author}
  {\bibfnamefont{J.}~\bibnamefont{Wu}},\ and\ \bibinfo {author}
  {\bibfnamefont{P.}~\bibnamefont{Phillips}},\ }%
  \bibfield{journal}{%
  \bibinfo {journal} {Phys. Rev. B}\ }%
  \textbf{\bibinfo {volume} {80}},\ \bibinfo {pages} {224506} (\bibinfo {year}
  {2009})%
  \bibAnnoteFile{NoStop}{Lv:2009p260}%
\bibitem{Lee:2009}%
  \BibitemOpen
  \bibfield{author}{%
  \bibinfo {author} {\bibfnamefont{C.-C.}\ \bibnamefont{Lee}}, \bibinfo
  {author} {\bibfnamefont{W.-G.}\ \bibnamefont{Yin}},\ and\ \bibinfo {author}
  {\bibfnamefont{W.}~\bibnamefont{Ku}},\ }%
  \bibfield{journal}{%
  \bibinfo {journal} {Phys. Rev. Lett.}\ }%
  \textbf{\bibinfo {volume} {103}},\ \bibinfo {pages} {267001} (\bibinfo {year}
  {2009})%
  \bibAnnoteFile{NoStop}{Lee:2009}%
\bibitem{Brouet:2008}%
  \BibitemOpen
  \bibfield{author}{%
  \bibinfo {author} {\bibfnamefont{V.}~\bibnamefont{Brouet}} \emph{et~al.},\ }%
  \bibfield{journal}{%
  \bibinfo {journal} {Phys. Rev. B}\ }%
  \textbf{\bibinfo {volume} {77}},\ \bibinfo {pages} {235104} (\bibinfo {year}
  {2008})%
  \bibAnnoteFile{NoStop}{Brouet:2008}%
\bibitem{Zhang:2009}%
  \BibitemOpen
  \bibfield{author}{%
  \bibinfo {author} {\bibfnamefont{Y.}~\bibnamefont{Zhang}} \emph{et~al.},\ }%
  \bibfield{journal}{%
  \bibinfo {journal} {Phys. Rev. B}\ }%
  \textbf{\bibinfo {volume} {83}},\ \bibinfo {pages} {054510} (\bibinfo {year}
  {2009})%
  \bibAnnoteFile{NoStop}{Zhang:2009}%
\bibitem{Mansart:2010}%
  \BibitemOpen
  \bibfield{author}{%
  \bibinfo {author} {\bibfnamefont{B.}~\bibnamefont{Mansart}} \emph{et~al.},\
  }%
  \bibfield{journal}{%
  \bibinfo {journal} {Phys. Rev. B}\ }%
  \textbf{\bibinfo {volume} {83}},\ \bibinfo {pages} {064516} (\bibinfo {year}
  {2011})%
  \bibAnnoteFile{NoStop}{Mansart:2010}%
\bibitem{Fink:2009p30}%
  \BibitemOpen
  \bibfield{author}{%
  \bibinfo {author} {\bibfnamefont{J.}~\bibnamefont{Fink}} \emph{et~al.},\ }%
  \bibfield{journal}{%
  \bibinfo {journal} {Phys. Rev. B}\ }%
  \textbf{\bibinfo {volume} {79}},\ \bibinfo {pages} {155118} (\bibinfo {year}
  {2009})%
  \bibAnnoteFile{NoStop}{Fink:2009p30}%
\bibitem{ZabolotnyyNature09}%
  \BibitemOpen
  \bibfield{author}{%
  \bibinfo {author} {\bibfnamefont{V.}~\bibnamefont{Zabolotnyy}}
  \emph{et~al.},\ }%
  \bibfield{journal}{%
  \bibinfo {journal} {Nature}\ }%
  \textbf{\bibinfo {volume} {457}},\ \bibinfo {pages} {569} (\bibinfo {year}
  {2009})%
  \bibAnnoteFile{NoStop}{ZabolotnyyNature09}%
\bibitem{Hsieh:2008p64}%
  \BibitemOpen
  \bibfield{author}{%
  \bibinfo {author} {\bibfnamefont{D.}~\bibnamefont{Hsieh}} \emph{et~al.},\ }%
  \bibinfo {journal} {arXiv:0812.2289v1}%
  \bibAnnoteFile{NoStop}{Hsieh:2008p64}%
\bibitem{Jong:2010p95}%
  \BibitemOpen
\bibfield{journal}{%
    }%
  \bibfield{author}{%
  \bibinfo {author} {\bibfnamefont{S.}~\bibnamefont{de~Jong}} \emph{et~al.},\
  }%
  \bibfield{journal}{%
  \bibinfo {journal} {Europhys. Lett.}\ }%
  \textbf{\bibinfo {volume} {89}},\ \bibinfo {pages} {27007} (\bibinfo {year}
  {2010})%
  \bibAnnoteFile{NoStop}{Jong:2010p95}%
\bibitem{LiuZhouPRB09}%
  \BibitemOpen
  \bibfield{author}{%
  \bibinfo {author} {\bibfnamefont{G.}~\bibnamefont{Liu}} \emph{et~al.},\ }%
  \bibfield{journal}{%
  \bibinfo {journal} {Phys. Rev. B}\ }%
  \textbf{\bibinfo {volume} {80}},\ \bibinfo {pages} {134519} (\bibinfo {year}
  {2009})%
  \bibAnnoteFile{NoStop}{LiuZhouPRB09}%
\bibitem{LiuKaminskiNaturePhys10}%
  \BibitemOpen
  \bibfield{author}{%
  \bibinfo {author} {\bibfnamefont{C.}~\bibnamefont{Liu}} \emph{et~al.},\ }%
  \bibfield{journal}{%
  \bibinfo {journal} {Nature Physics}\ }%
  \textbf{\bibinfo {volume} {6}},\ \bibinfo {pages} {419} (\bibinfo {year}
  {2010})%
  \bibAnnoteFile{NoStop}{LiuKaminskiNaturePhys10}%
\bibitem{Richard:2010p62}%
  \BibitemOpen
  \bibfield{author}{%
  \bibinfo {author} {\bibfnamefont{P.}~\bibnamefont{Richard}} \emph{et~al.},\
  }%
  \bibfield{journal}{%
  \bibinfo {journal} {Phys. Rev. Lett.}\ }%
  \textbf{\bibinfo {volume} {104}},\ \bibinfo {pages} {137001} (\bibinfo {year}
  {2010})%
  \bibAnnoteFile{NoStop}{Richard:2010p62}%
\bibitem{Kondo:2010p261}%
  \BibitemOpen
  \bibfield{author}{%
  \bibinfo {author} {\bibfnamefont{T.}~\bibnamefont{Kondo}} \emph{et~al.},\ }%
  \bibfield{journal}{%
  \bibinfo {journal} {Phys. Rev. B}\ }%
  \textbf{\bibinfo {volume} {81}},\ \bibinfo {pages} {060507} (\bibinfo {year}
  {2010})%
  \bibAnnoteFile{NoStop}{Kondo:2010p261}%
\bibitem{Yang:2009p262}%
  \BibitemOpen
  \bibfield{author}{%
  \bibinfo {author} {\bibfnamefont{L.~X.}\ \bibnamefont{Yang}} \emph{et~al.},\
  }%
  \bibfield{journal}{%
  \bibinfo {journal} {Phys. Rev. Lett.}\ }%
  \textbf{\bibinfo {volume} {102}},\ \bibinfo {pages} {107002} (\bibinfo {year}
  {2009})%
  \bibAnnoteFile{NoStop}{Yang:2009p262}%
\bibitem{Yi:2009p234}%
  \BibitemOpen
  \bibfield{author}{%
  \bibinfo {author} {\bibfnamefont{M.}~\bibnamefont{Yi}} \emph{et~al.},\ }%
  \bibfield{journal}{%
  \bibinfo {journal} {Phys. Rev. B}\ }%
  \textbf{\bibinfo {volume} {80}},\ \bibinfo {pages} {174510} (\bibinfo {year}
  {2009})%
  \bibAnnoteFile{NoStop}{Yi:2009p234}%
\bibitem{KimPRB11}%
  \BibitemOpen
  \bibfield{author}{%
  \bibinfo {author} {\bibfnamefont{Y.}~\bibnamefont{Kim}} \emph{et~al.},\ }%
  \bibfield{journal}{%
  \bibinfo {journal} {Phys. Rev. B}\ }%
  \textbf{\bibinfo {volume} {83}},\ \bibinfo {pages} {064509} (\bibinfo {year}
  {2011})%
  \bibAnnoteFile{NoStop}{KimPRB11}%
\bibitem{YiPNAS11}%
  \BibitemOpen
  \bibfield{author}{%
  \bibinfo {author} {\bibfnamefont{M.}~\bibnamefont{Yi}} \emph{et~al.},\ }%
  \bibfield{journal}{%
  \bibinfo {journal} {PNAS}\ }%
  \textbf{\bibinfo {volume} {108}},\ \bibinfo {pages} {6878} (\bibinfo {year}
  {2011})%
  \bibAnnoteFile{NoStop}{YiPNAS11}%
\bibitem{DessauUntwinned}%
  \BibitemOpen
  \bibfield{author}{%
  \bibinfo {author} {\bibfnamefont{Q.}~\bibnamefont{Wang}} \emph{et~al.},\ }%
  \bibinfo {journal} {arXiv:1009.0271v1}%
  \bibAnnoteFile{NoStop}{DessauUntwinned}%
\bibitem{Shimojima:2010p37}%
  \BibitemOpen
\bibfield{journal}{%
    }%
  \bibfield{author}{%
  \bibinfo {author} {\bibfnamefont{T.}~\bibnamefont{Shimojima}} \emph{et~al.},\
  }%
  \bibfield{journal}{%
  \bibinfo {journal} {Phys. Rev. Lett.}\ }%
  \textbf{\bibinfo {volume} {104}},\ \bibinfo {pages} {057002} (\bibinfo {year}
  {2010})%
  \bibAnnoteFile{NoStop}{Shimojima:2010p37}%
\bibitem{BrouetJensen2011}%
  \BibitemOpen
  \bibfield{author}{%
  \bibinfo {author} {\bibfnamefont{V.}~\bibnamefont{Brouet}} \emph{et~al.},\ }%
  \bibfield{journal}{%
  \bibinfo {journal} {available on cond-mat arxiv}}%
   (\bibinfo {year} {2011})%
  \bibAnnoteFile{NoStop}{BrouetJensen2011}%
\bibitem{Brouet:2009p106}%
  \BibitemOpen
  \bibfield{author}{%
  \bibinfo {author} {\bibfnamefont{V.}~\bibnamefont{Brouet}} \emph{et~al.},\ }%
  \bibfield{journal}{%
  \bibinfo {journal} {Phys Rev B}\ }%
  \textbf{\bibinfo {volume} {80}},\ \bibinfo {pages} {165115} (\bibinfo {year}
  {2009})%
  \bibAnnoteFile{NoStop}{Brouet:2009p106}%
\bibitem{Graser:2010p67}%
  \BibitemOpen
  \bibfield{author}{%
  \bibinfo {author} {\bibfnamefont{S.}~\bibnamefont{Graser}} \emph{et~al.},\ }%
  \bibfield{journal}{%
  \bibinfo {journal} {Phys. Rev. B}\ }%
  \textbf{\bibinfo {volume} {81}},\ \bibinfo {pages} {214503} (\bibinfo {year}
  {2010})%
  \bibAnnoteFile{NoStop}{Graser:2010p67}%
\bibitem{ZhouFengPRB10}%
  \BibitemOpen
  \bibfield{author}{%
  \bibinfo {author} {\bibfnamefont{B.}~\bibnamefont{Zhou}} \emph{et~al.},\ }%
  \bibfield{journal}{%
  \bibinfo {journal} {Phys. Rev. B}\ }%
  \textbf{\bibinfo {volume} {81}},\ \bibinfo {pages} {155124} (\bibinfo {year}
  {2010})%
  \bibAnnoteFile{NoStop}{ZhouFengPRB10}%
\bibitem{Ran:2009p157}%
  \BibitemOpen
  \bibfield{author}{%
  \bibinfo {author} {\bibfnamefont{Y.}~\bibnamefont{Ran}} \emph{et~al.},\ }%
  \bibfield{journal}{%
  \bibinfo {journal} {Phys. Rev. B}\ }%
  \textbf{\bibinfo {volume} {79}} (\bibinfo {year} {2009})%
  \bibAnnoteFile{NoStop}{Ran:2009p157}%
\bibitem{moon:2010}%
  \BibitemOpen
  \bibfield{author}{%
  \bibinfo {author} {\bibfnamefont{S.~J.}\ \bibnamefont{Moon}} \emph{et~al.},\
  }%
  \bibfield{journal}{%
  \bibinfo {journal} {Phys. Rev. B}\ }%
  \textbf{\bibinfo {volume} {81}},\ \bibinfo {pages} {205114} (\bibinfo {year}
  {2010})%
  \bibAnnoteFile{NoStop}{moon:2010}%
\bibitem{Massee:2009p247}%
  \BibitemOpen
  \bibfield{author}{%
  \bibinfo {author} {\bibfnamefont{F.}~\bibnamefont{Massee}} \emph{et~al.},\ }%
  \bibfield{journal}{%
  \bibinfo {journal} {Phys. Rev. B}\ }%
  \textbf{\bibinfo {volume} {80}},\ \bibinfo {pages} {140507} (\bibinfo {year}
  {2009})%
  \bibAnnoteFile{NoStop}{Massee:2009p247}%
\bibitem{He:2010p245}%
  \BibitemOpen
  \bibfield{author}{%
  \bibinfo {author} {\bibfnamefont{C.}~\bibnamefont{He}} \emph{et~al.},\ }%
  \bibfield{journal}{%
  \bibinfo {journal} {Phys Rev Lett}\ }%
  \textbf{\bibinfo {volume} {105}},\ \bibinfo {pages} {117002} (\bibinfo {year}
  {2010})%
  \bibAnnoteFile{NoStop}{He:2010p245}%
\end{thebibliography}%

\end{document}